\documentclass{elsart}

 \usepackage{epsfig}

\usepackage{amssymb}

\begin{document}

\begin{frontmatter}



\title{Scenario of inflationary cosmology from the
phenomenological $\Lambda$ models}

\author[label1]{Saibal Ray}\footnote{Corresponding author (E-mail: saibal@iucaa.ernet.in).},
\author[label2]{Pratap Chandra Ray},
\author[label3,label4,label5]{Maxim Khlopov},
\author[label6]{Partha Pratim Ghosh},
\author[label7]{Utpal Mukhopadhyay},
\author[label8]{Partha Chowdhury}
\address[label1]{Department of Physics, Barasat Government College,
Barasat 700124, North 24 Parganas, West Bengal, India}
\address[label2]{Department of Mathematics, Government College of
Engineering and Leather Technology, Kolkata 700 098, West Bengal,
India}
\address[label3]{Center for Cosmoparticle physics ``Cosmion",
125047, Moscow, Russia}
\address[label4]{Moscow Engineering Physics Institute, 115409 Moscow, Russia}
\address[label5]{APC laboratory 10, rue Alice Domon et Léonie Duquet 75205 Paris
Cedex 13, France}
\address[label6]{Tara Brahmamoyee Vidyamandir, Matripalli, Shyamnagar 743
127, North 24 Parganas, West Bengal, India}
\address[label7]{Satyabharati Vidyapith, Nabapalli, North 24 Parganas,
Kolkata 700 126, West Bengal, India.}
\address[label8]{International Centre for Complex System Studies, West
Bengal University of Technology, Kolkata 700 064, West Bengal,
India}

\begin{abstract}
 Choosing the three phenomenological models of the
dynamical cosmological term $\Lambda$, viz., $\Lambda \sim (\dot
a/a)^2$,
 $\Lambda \sim {\ddot a/a}$ and $\Lambda \sim \rho$ where $a$
 is the cosmic scale factor, it has been shown by the method of
 numerical analysis for the considered non-linear differential equations
 that the three models are equivalent for the
 flat Universe $k=0$ and for arbitrary non-linear equation of state.
 The evolution plots for dynamical
cosmological term $\Lambda$ vs. time $t$ and also the cosmic scale
factor $a$ vs. $t$ are drawn here for $k=0, +1$. A qualitative
analysis has been made from the plots which supports the idea of
inflation and hence expanding Universe.
\end{abstract}

\end{frontmatter}

\section{Introduction}
The observations on Supernovae type
Ia~\cite{Riess1998,Perlmutter1998} lead the scientists to the
concept of accelerating Universe. The unknown
 type of energy responsible for this kind of acceleration is known as
 {\it dark energy}. Now, there are various types of models related to this
 dark energy~\cite{Overduin1998,Sahni2000} and for those models the expansion rate
 is also different. The so called {\it cosmological constant} may be one of them. For the
 explanation of accelerating Universe, $\Lambda$ of dynamical character is preferred
 rather than a constant one. This character can naturally follow from non-linear effects of
 gravitational and scalar fields, involved in the physical model. In the paper of
 Ray, Mukhopadhyay and Meng ~\cite{Ray2007}, it was shown that among the dynamical
 models of $\Lambda$, the three types $\Lambda \sim (\dot a/a)^2$,
 $\Lambda \sim {\ddot a/a}$ and $\Lambda \sim \rho$ are
 equivalent, where $a$ is the cosmic scale factor of the Robertson-Walker
 metric and $\rho$ is the matter-energy density. They also
 analytically established a relationship between the parameters
 $\alpha$, $\beta$ and $\gamma$ of the respective models. In the
 present work emphasis has been given to show the equivalence of the
 same three models of $\Lambda$ by using the method of numerical
 analysis of underlying non-linear Einstein equation for
 the flat Universe ($k=0$) and for equation of state with
 arbitrary non-linear dependence on matter-energy density.
 Another aspect of this work is to make an attempt for
 visualizing the incidents that occurred during the very early stage
 of the Universe, especially the feature of {\it inflation}
 ~\cite{Guth1981,Linde1982,Albrecht1982}.

\section{The Field Equations and General Results}
The Einstein field equations are given by
 \begin{eqnarray}
  R^{ij} - \frac{1}{2}Rg^{ij} = -8 \pi G \left[T^{ij} - \frac{\Lambda}{8\pi G}g^{ij}\right]
\end{eqnarray}
 where $\Lambda$= $\Lambda(t)$ is the so called {\it cosmological
 constant}. Here $c$, the velocity of light, is assumed to be unity in relativistic units.

 Now, let us consider the Friedmann-Lema{\^i}tre-Robertson-Walker
 metric
 \begin{eqnarray}
ds^2 = -dt^2 + a(t)^2\left[\frac{dr^2}{1 - kr^2} + r^2 (d\theta^2
+ sin^2\theta d\phi^2)\right]
\end{eqnarray}
where the curvature constant $k=-1, 0, +1$ for open, flat and
close models of the Universe respectively.

 For the above spherically
symmetric metric (2), the non-linear Einstein field equations
reduce to the following two equations, respectively the Friedmann
equation and the Raychaudhuri equation as
\begin{eqnarray}
\left(\frac{\dot a}{a}\right)^2 + \frac{k}{a^2}= \frac{8 \pi G
\rho}{3}+ \frac{\Lambda}{3},
\end{eqnarray}
\begin{eqnarray}
\frac{\ddot a}{a}=-\frac{4\pi G}{3}\left(\rho + 3p \right) +
\frac{\Lambda}{3}
\end{eqnarray}
where the fluid pressure $p$ and energy density $\rho$ are related
in the form
\begin{eqnarray}
p= \omega \rho^n
\end{eqnarray}
$\omega$ being the equation of state parameter of the polytropic
equation of state.

The energy conservation law can be written as
\begin{eqnarray}
8\pi G(p+\rho)\frac{\dot a}{a}= -\frac{8\pi G}{3}
\dot\rho-\frac{\dot\Lambda}{3}.
\end{eqnarray}
Differentiating equation (3) with respect to time $t$ we get
\begin{eqnarray}
2\left( \frac{\dot a}{a}\right)\left[\frac{a\ddot a-\dot
a^2}{a^2}\right] -\frac{2k}{a^3} \dot a= \frac{8\pi G}{3}\dot \rho
+ \frac{\dot \Lambda}{3}.
\end{eqnarray}
With the help of equation (6) equation (7) reduces to
\begin{eqnarray}
 \left( \frac{\dot a}{a}\right)^2 + \frac{k}{a^2} -\frac{\ddot a}{a}= 4\pi G
 (p+\rho).
\end{eqnarray}
Now, from equation (4) with the help of equations (8) and (5), we
get \begin{eqnarray} \left( \frac{\dot a}{a}\right)^2
+2\left(\frac{\ddot a}{a}\right)+ \frac {k}{a^2} =-8\pi G\omega
\rho^n+\Lambda.
\end{eqnarray}

\subsection{$\Lambda \sim (\dot
a/a)^2$} Let us now consider the following dynamical model of
$\Lambda$ which is
\begin{eqnarray} \Lambda= 3\alpha\left( \frac{\dot
a}{a}\right)^2.
\end{eqnarray}
Employing this equation (10) in equation (3) one immediately
obtains
\begin{eqnarray}
\rho= \frac{3}{8\pi G}\left[\left( \frac{\dot
a}{a}\right)^2+\frac{k}{a^2}\right]-\frac{3\alpha}{8\pi G}\left(
\frac{\dot a}{a}\right)^2.
\end{eqnarray}

Now taking $n=1$ and using equation (11), equation (9) reduces to
the form
\begin{eqnarray}
\ddot a=-(3\omega+1)\left[\frac{\dot
a^2}{2a}+\frac{k}{a}\right]+3\alpha(1+\omega)\frac{\dot a^2}{2a}.
\end{eqnarray}

\subsection{$\Lambda \sim {\ddot a/a}$} As a second dynamical model we now start with
\begin{eqnarray} \Lambda= \beta\left(\frac{\ddot a}{a}\right).
\end{eqnarray}
By the use of equation (13) in equation (3), we get the value for
$\rho$ as
\begin{eqnarray}
\rho= \frac{3}{8\pi G}\left[\left( \frac{\dot
a}{a}\right)^2+\frac{k}{a^2}\right]-\frac{\beta}{8 \pi
G}\left(\frac{\ddot a}{a}\right).
\end{eqnarray}
For $n=1$, by the use of equation (14), the equation (9) finally
gives the value of $\ddot a$ as
\begin{eqnarray}
\ddot a=\frac{(1+3\omega)}{[(\omega+1)\beta-2]}\left[\frac{\dot
a^2}{a}+\frac{k}{a}\right].
\end{eqnarray}

\subsection{$\Lambda \sim \rho$} Let us now take the third form of phenomenological
$\Lambda$ as
\begin{eqnarray}
\Lambda=8\pi G \gamma \rho
\end{eqnarray} where the constraint is such that $\gamma>0$.
Using the value of $\Lambda$ given in equation (16), we obtain
from equation (3) the value of $\rho$ as
\begin{eqnarray}
\rho=\frac{3}{8 \pi G(1+\gamma)}\left[\left( \frac{\dot
a}{a}\right)^2+ \frac{k}{a^2}\right].
\end{eqnarray}
Using equation (16) and (17), we get from equation (9) the value
of $\ddot a$ for $n=1$ as
\begin{eqnarray}
\ddot a=\frac{2\gamma-3\omega-1}{2(\gamma+1)}\left[\frac{\dot
a^2}{a}+\frac{k}{a}\right].
\end{eqnarray}

 In this connection it is to be noted here that Ray, Mukhopadhyay and
 Meng~\cite{Ray2007} have shown that the three forms $\Lambda = 3\alpha (\dot a/a)^2$,
 $\Lambda = \beta (\ddot a/a)$ and $\Lambda = 8\pi G \gamma \rho$,
 as expressed in equations (10), (13) and (16), are
 equivalent for $k=0$. They found through analytical method that the
 parameters involved in the
 three dynamical relations are connected by
\begin{eqnarray}
    \alpha = \frac{\beta(1 + 3w)}{3(\beta w + \beta - 2)}
            = \frac{\gamma}{1 + \gamma}.
\end{eqnarray}
  This means that it is possible to find out the
 identical physical features of others if any of those three phenomenological
 $\Lambda$ relations is known.

In this connection it is, however, interesting to note here that
for linear dependence on energy density in equation of state (5),
corresponding to $n=1$, the equation (15) of {\it Case-2.2} and
the equation (18) of {\it Case-2.3} are equivalent for $k=+1, 0,
-1$ when the relation
 \begin{eqnarray}
\frac{(1+3\omega)}{[(\omega+1)\beta-2]}=\frac{2\gamma-3\omega-1}{2(\gamma+1)}
\end{eqnarray}
holds good.

\section{Graphical Presentation of the Results}
Now using the method of numerical analysis, let us try to study
the variation of the cosmological parameter $\Lambda$ and the
scale factor $a$ with time which are shown in the following plots.

From a close observation of the graphical plots, viz.,
 figures 1(a), 2(a) and 3(a), one can find out equivalence
 of the three models with respect to time variation of
 $\Lambda$ while figures 1(b), 2(b) and 3(b) exhibit
 equivalence of the same three models with respect to time variation of
 $a$ for $k=0$ as obtained in analytical method by
 Ray, Mukhopadhyay and Meng~\cite{Ray2007}. It can also be observed
 that figures 2(c) and 3(c) show
 equivalence of $\Lambda \sim {\ddot a/a}$ and $\Lambda \sim \rho$
 models with respect to variation of $\Lambda$ with time for
 $k=+1$. However, the behaviour of $\Lambda$ with respect to time
 for $\Lambda \sim (\dot a/a)^2$ model is quite different. For
 $\Lambda \sim (\dot a/a)^2$ model we observe an abrupt increase
 of the cosmological parameter within a very short period of time
 and then a comparatively slower decrease of it. This abrupt rise
 of $\Lambda$ may be interpreted as the driving force behind
 inflation because time variation of the scale factor $a$ for the same
 values of $k$, $\omega$, $\alpha$, $\beta$ and $\gamma$ show a
 very sharp increase as depicted in figures 1(d), 2(d) and
 3(d). Moreover, the sudden jump in the value of $\Lambda$
 exhibited in figure 1(c) for $k=1$ shows a clear
indication of the role of the dark energy candidate $\Lambda$ as
repulsive pressure in connection to inflationary phase of the
Universe ~\cite{Guth1981,Linde1982,Albrecht1982} and also may be
interpreted as a numerical manifestation of the idea that dark
energy is responsible for making the Universe flat during
inflation. This is because immediately after attaining a peak
value, $\Lambda$ has dropped down. So, it is quite natural to
think of that a huge amount of dark energy was used up for
triggering the exponential cosmic expansion as well as for
removing the curvature of the Universe. Interestingly, it is to be
noted here that this type of qualitative variation of $\Lambda$
with time during inflation was unavailable in the analytical
 method by Ray, Mukhopadhyay and Meng~\cite{Ray2007}.
 However, due to lack of advanced
computing facility, in the present investigation we could not
quantify the time duration of inflation which has been suggested
in the literature as $10^{-35}$ to $10^{-32}$ sec~\cite{Guth1998}.

The present work shows that although phenomenological models do
not originate from any quantum field theory, yet at least in some
cases they can successfully reflect the present cosmological
picture. This is another positive side of this investigation. In
particular taking the values of $\Omega_{m0}= 0.330 \pm
0.035.$~\cite{Riess1998,Perlmutter1998,Overduin1998,Sahni2000},
the ranges of the values of model parameters $\alpha_0, \beta_0,
\gamma_0$, corresponding to the observed accelerated expansion,
are obtained as $0.635 \ge \alpha_0 \ge 0.705, 3.417 \ge \beta_0
\ge 4.674$ and $1.739 \ge \gamma_0 \ge 2.389$.

Using the above-mentioned values of $\Omega_{m0}$, the ranges of
the present values of the cosmological parameter $\Lambda_0$ are
obtained as $1 \times 10^{-35}$ s$^{-2}$ - $2 \times 10^{-35}$
s$^{-2}$, which agree with the results of Carmeli~\cite{carmeli02}
and Carmeli and Kuzmenko~ \cite{car02k}, where they obtain the
value $1.934 \times 10^{-35}$ s$^{-2}$.

\section{Conclusions}
In the present investigation, instead of finding out exact
solution of the ordinary non-linear differential equation, the
method of numerical analysis has been adopted for the three
phenomenological models of $\Lambda$, viz., $\Lambda \sim (\dot
a/a)^2$, $\Lambda \sim {\ddot a/a}$ and $\Lambda \sim \rho$.
However, the main idea of the article i.e. the study of numerical
solutions for cosmological problems in absence of particular or
general analytic solutions is not a new idea. Interesting
literature in this aspect are available, where numerical cosmology
with the Cactus code~\cite{Vulcanov2002a} or testing the Cactus
code on exact solutions of the Einstein field
equations~\cite{Vulcanov2002b} have been performed.

The dynamical nature of $\Lambda$ and its time variation of the
present investigation can find physical origin in non-linear
effects of gravity (as it was the case for nonlinear $R^2$ term
induced by vacuum polarization in one of the first inflationary
models \cite{Starobinsky:1980te}) or of scalar fields, involved in
the model (as it took place in the of self consistent inflation
\cite{Dymnikova:2001ga,Dymnikova:2001jy,Dymnikova:1998ps}). The
interesting features of this investigation can be put in the
following way:\\ (1) The time variation of $\Lambda$ and $a$ for
$k=0$ support the work of Ray, Mukhopadhyay and
Meng~\cite{Ray2007} so far as the equivalence of the three chosen
phenomenological models are concerned; \\(2) It has been possible
to establish that $\Lambda \sim {\ddot a/a}$ and $\Lambda \sim
\rho$ models are also equivalent for $k=1$; \\(3) Finally, a bonus
obtained from the present numerical work is the qualitative
visualization of inflationary scenario of the Universe through the
variation of $\Lambda$.

\section*{Acknowledgments}
One of the authors (SR) is thankful to the authority of
Inter-University Centre for Astronomy and Astrophysics, Pune,
India, for providing Associateship programme under which a part of
this work was carried out. We all are also thankful to the referee
for valuable suggestions which have enabled us to improve the
manuscript substantially.

\begin{figure*}
\begin{center}
\vspace{0.5cm} \psfig{file=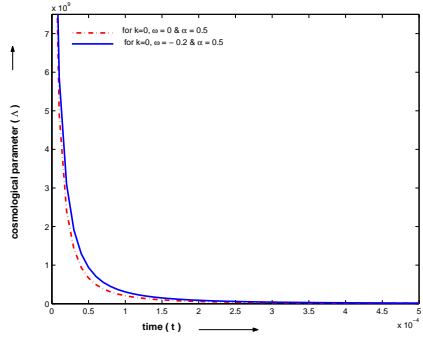,width=0.4\textwidth}
    \caption{Case-2.1 related to model $\Lambda = 3\alpha (\dot a/a)^2$
    shows variation of cosmological parameter for $k=0$.}
 \label{fig:1a}
\end{center}
\end{figure*}

\begin{figure*}
\begin{center}
\vspace{0.5cm} \psfig{file=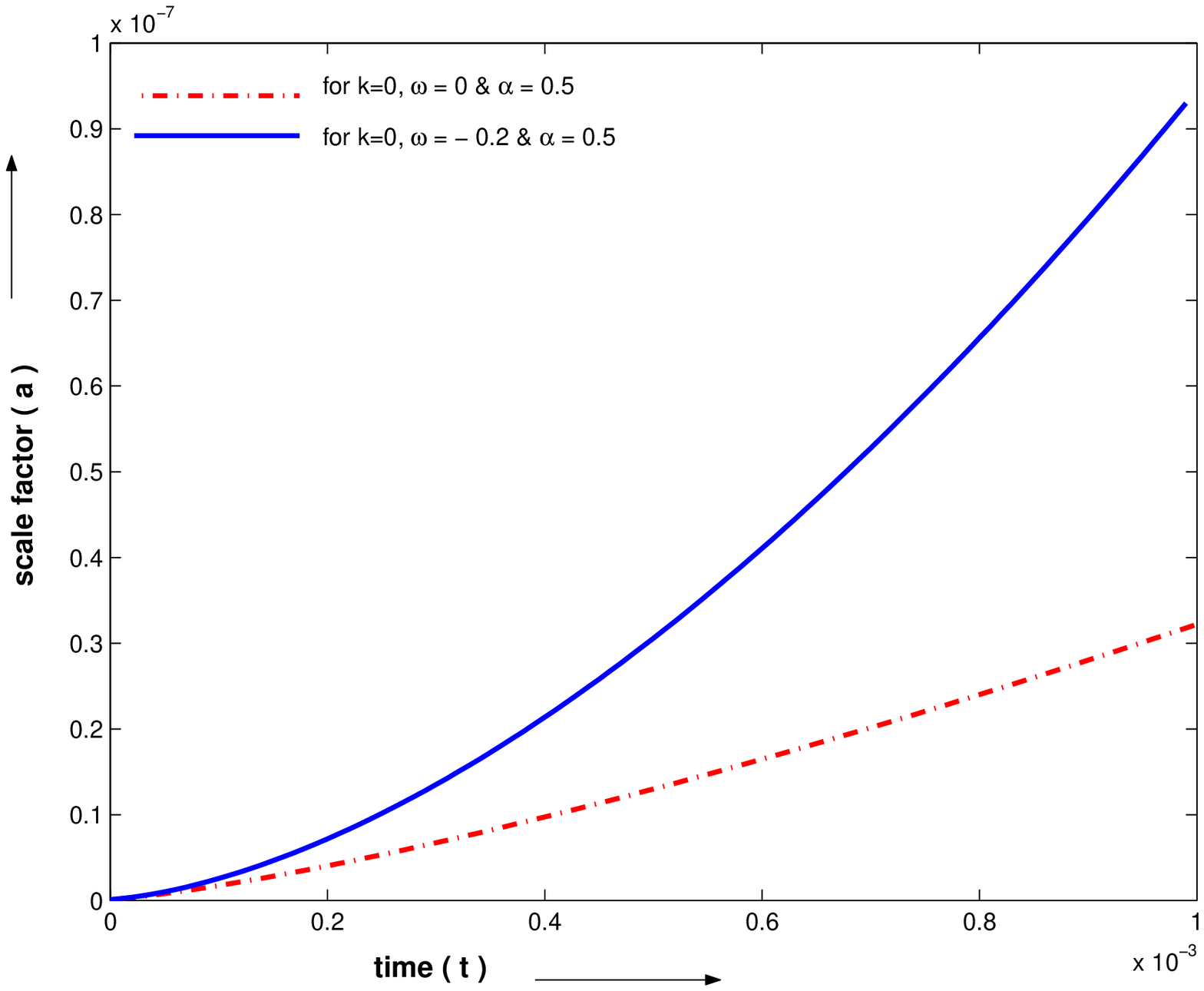,width=0.4\textwidth}
    \caption{Case-2.1 related to model $\Lambda = 3\alpha (\dot a/a)^2$
    shows variation of scale factor for $k=0$.}
 \label{fig:1b}
\end{center}
\end{figure*}

\begin{figure*}
\begin{center}
\vspace{0.5cm} \psfig{file=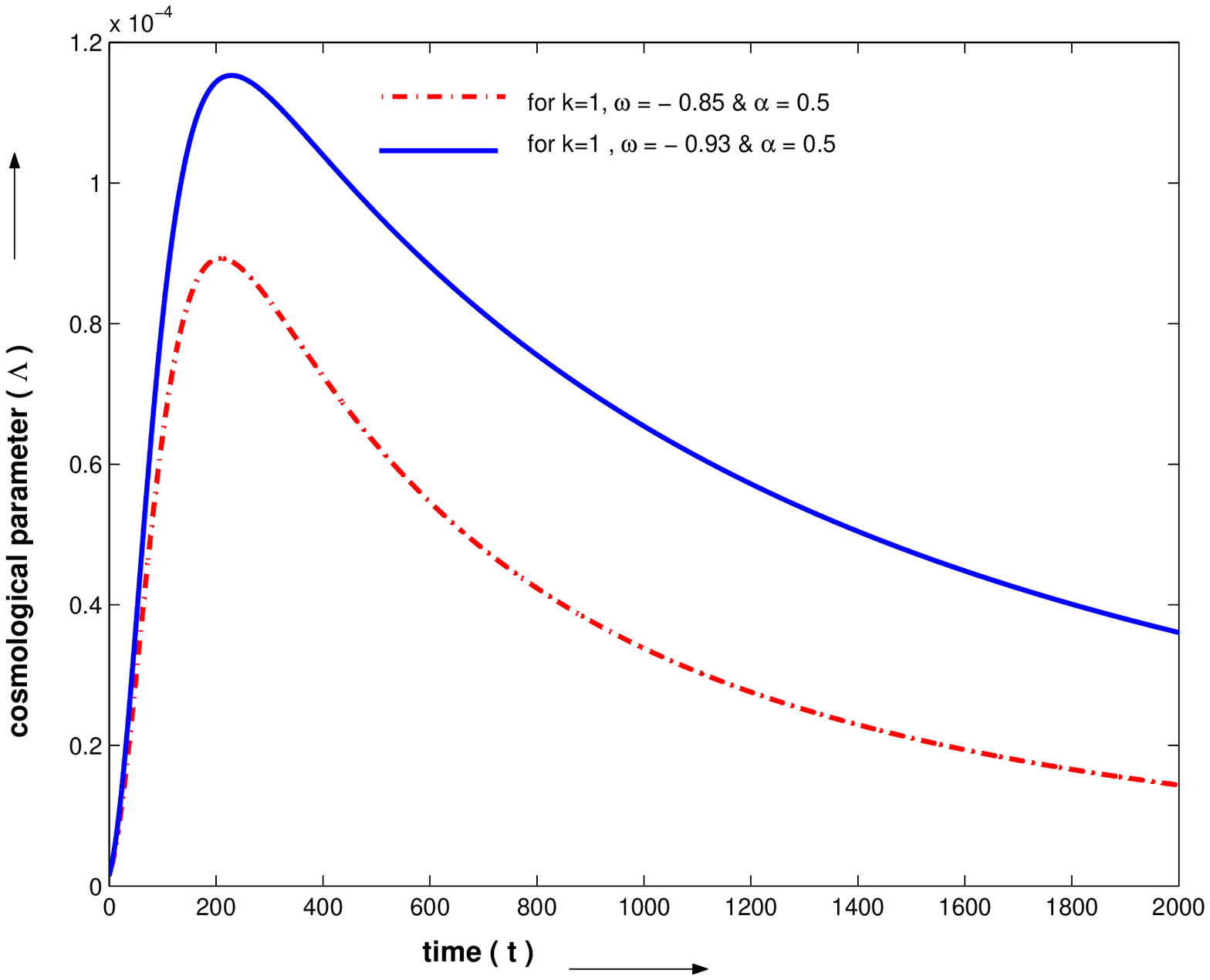,width=0.4\textwidth}
    \caption{Case-2.1 related to model $\Lambda = 3\alpha (\dot a/a)^2$
    shows variation of cosmological parameter for $k=+1$.}
 \label{fig:1c}
\end{center}
\end{figure*}

\begin{figure*}
\begin{center}
\vspace{0.5cm} \psfig{file=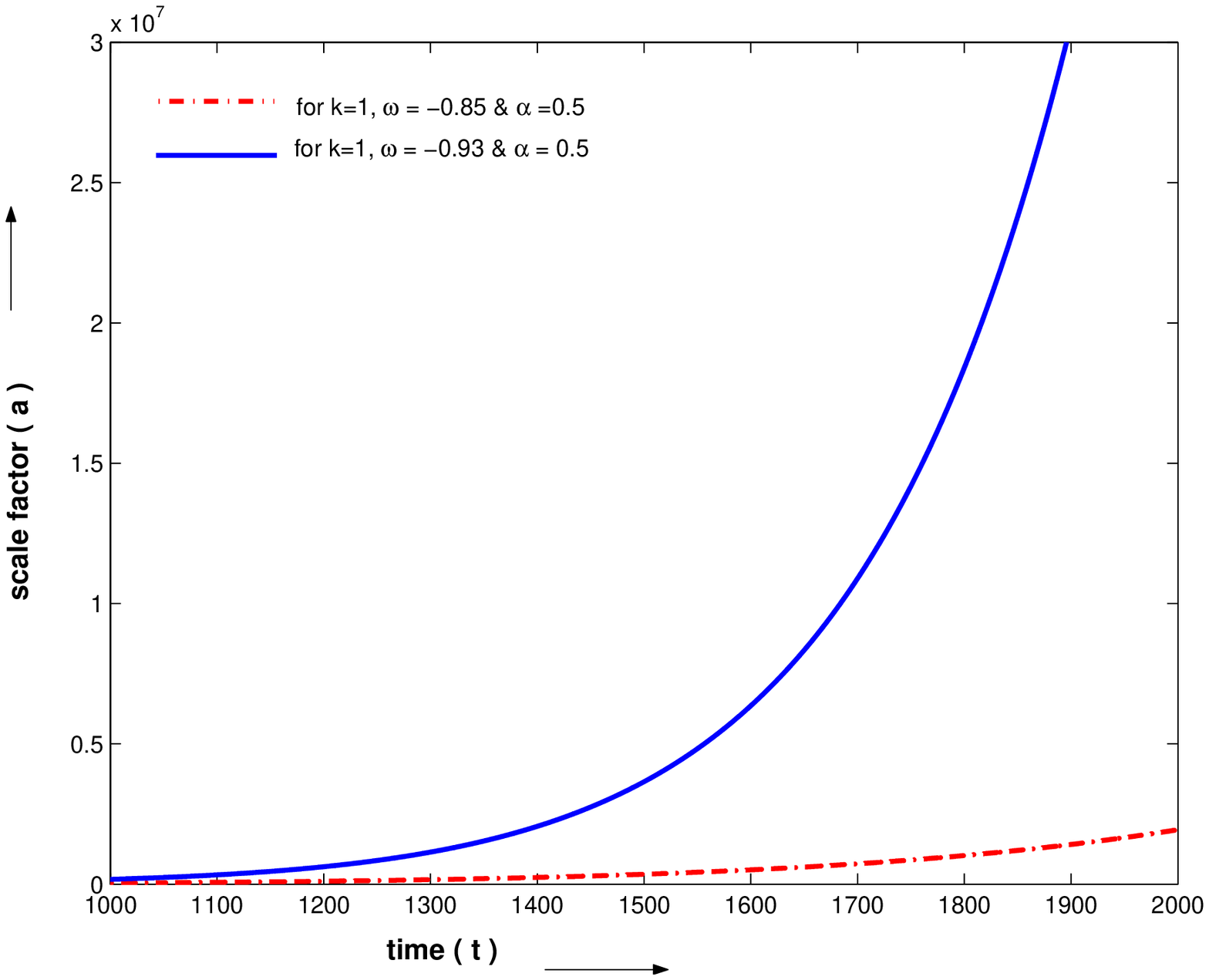,width=0.4\textwidth}
    \caption{Case-2.1 related to model $\Lambda = 3\alpha (\dot a/a)^2$
    shows variation of scale factor for $k=+1$.}
 \label{fig:1d}
\end{center}
\end{figure*}

\begin{figure*}
\begin{center}
\vspace{0.5cm} \psfig{file=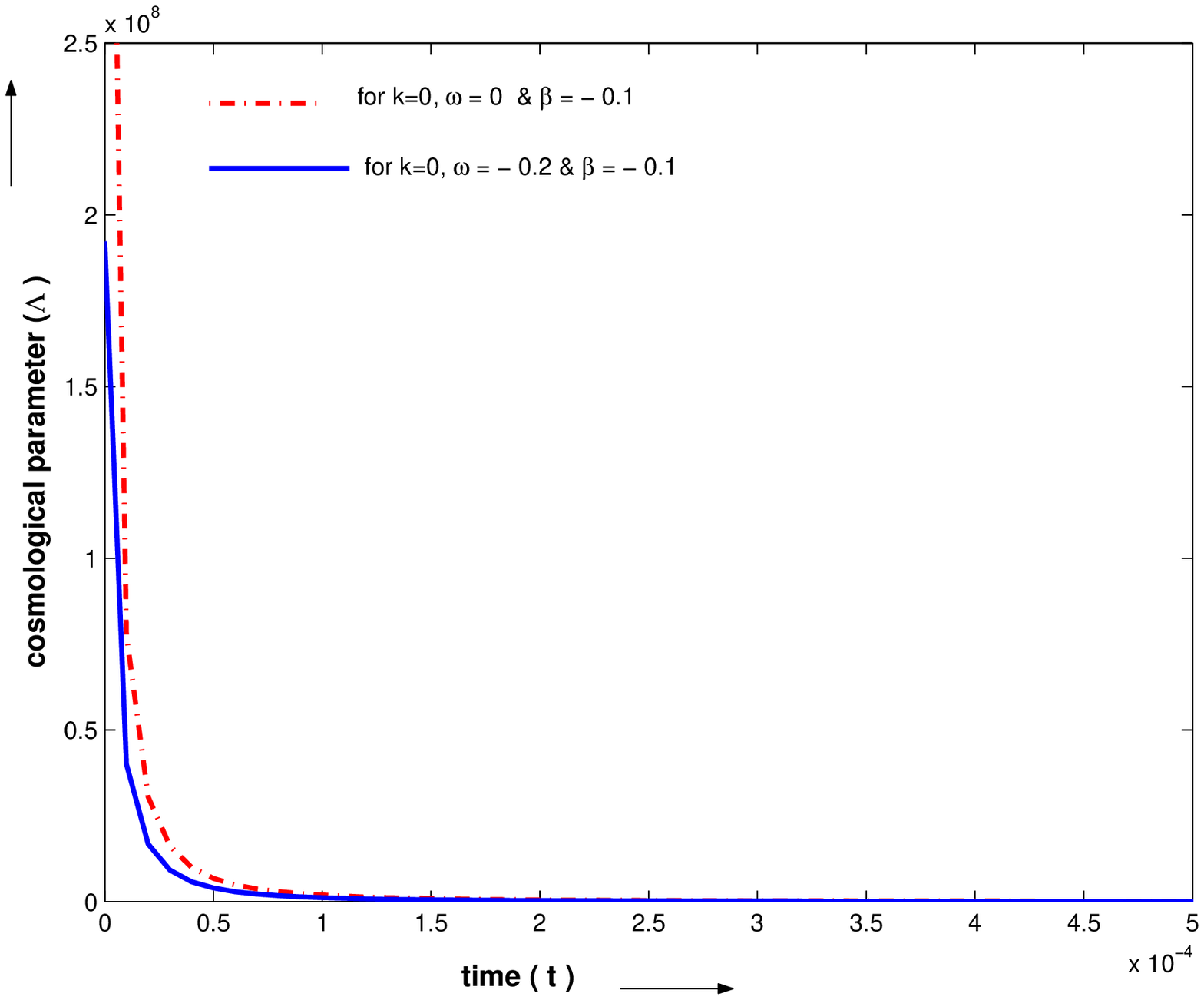,width=0.4\textwidth}
    \caption{Case-2.2 related to model $\Lambda = \beta (\ddot a/a)$
    shows variation of cosmological parameter for $k=0$.}
 \label{fig:2a}
\end{center}
\end{figure*}

\begin{figure*}
\begin{center}
\vspace{0.5cm} \psfig{file=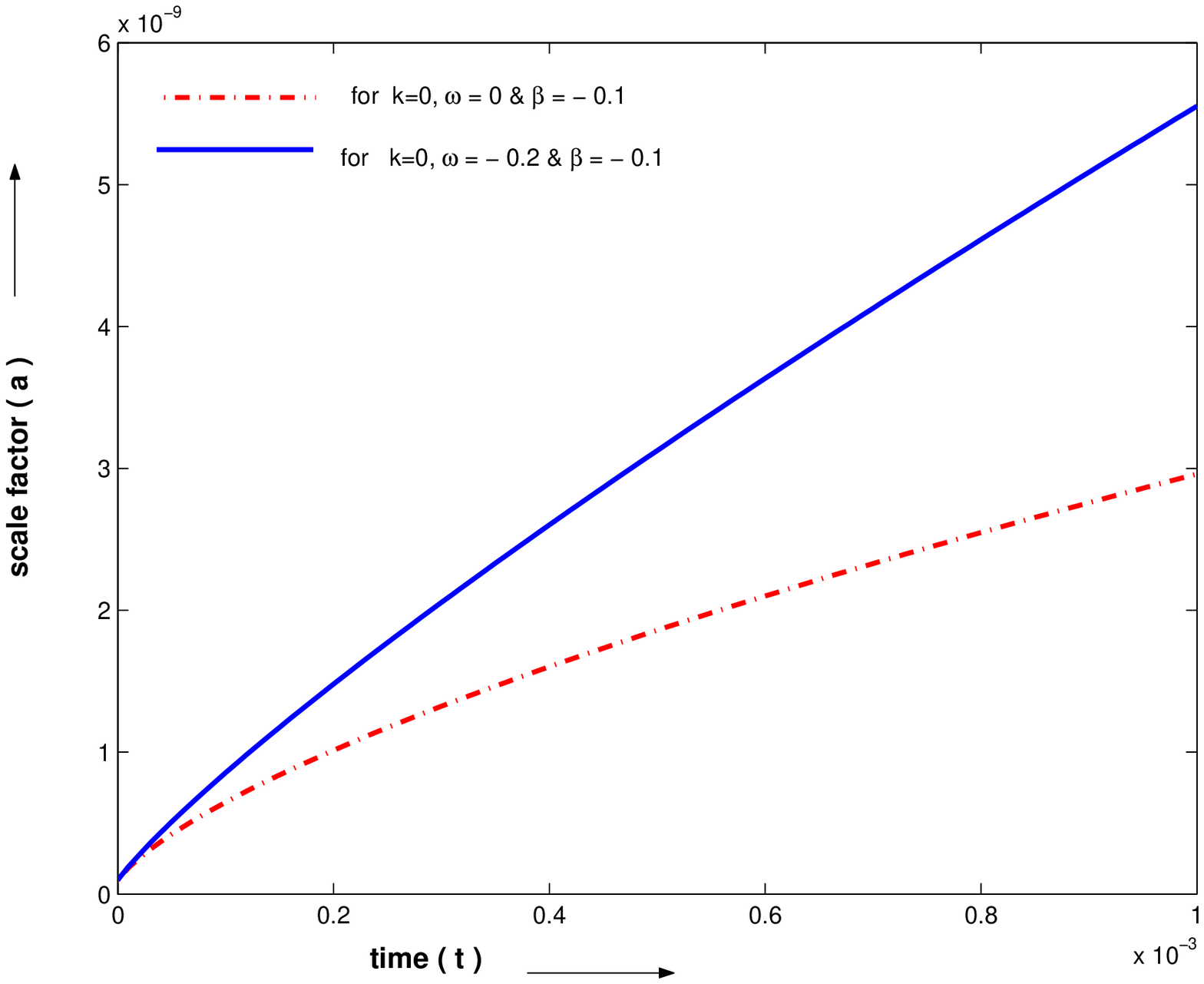,width=0.4\textwidth}
    \caption{Case-2.2 related to model $\Lambda = \beta (\ddot a/a)$
    shows variation of scale factor for $k=0$.}
 \label{fig:2b}
\end{center}
\end{figure*}

\begin{figure*}
\begin{center}
\vspace{0.5cm} \psfig{file=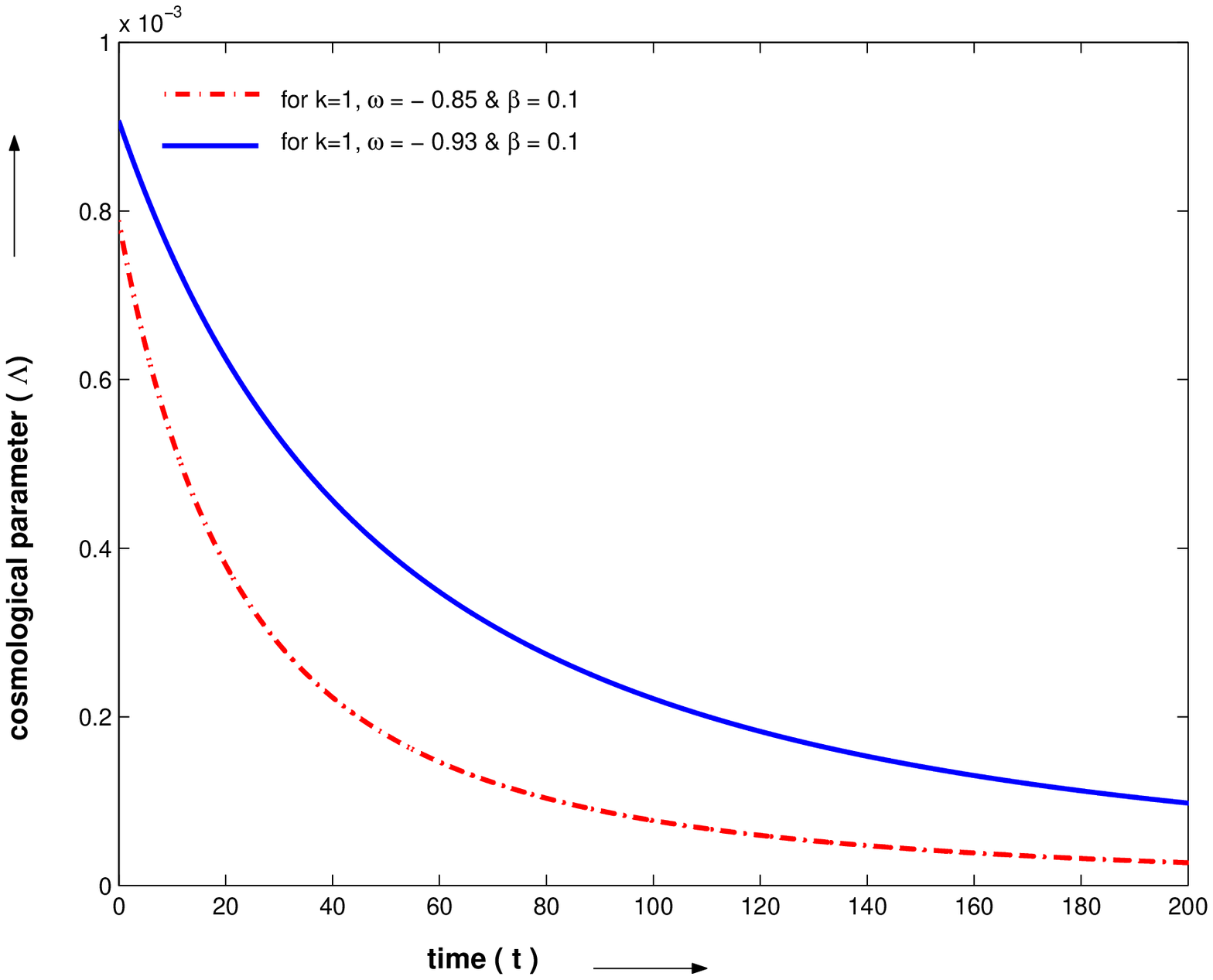,width=0.4\textwidth}
    \caption{Case-2.2 related to model $\Lambda = \beta (\ddot a/a)$
    shows variation of cosmological parameter for $k=+1$.}
 \label{fig:2c}
\end{center}
\end{figure*}

\begin{figure*}
\begin{center}
\vspace{0.5cm} \psfig{file=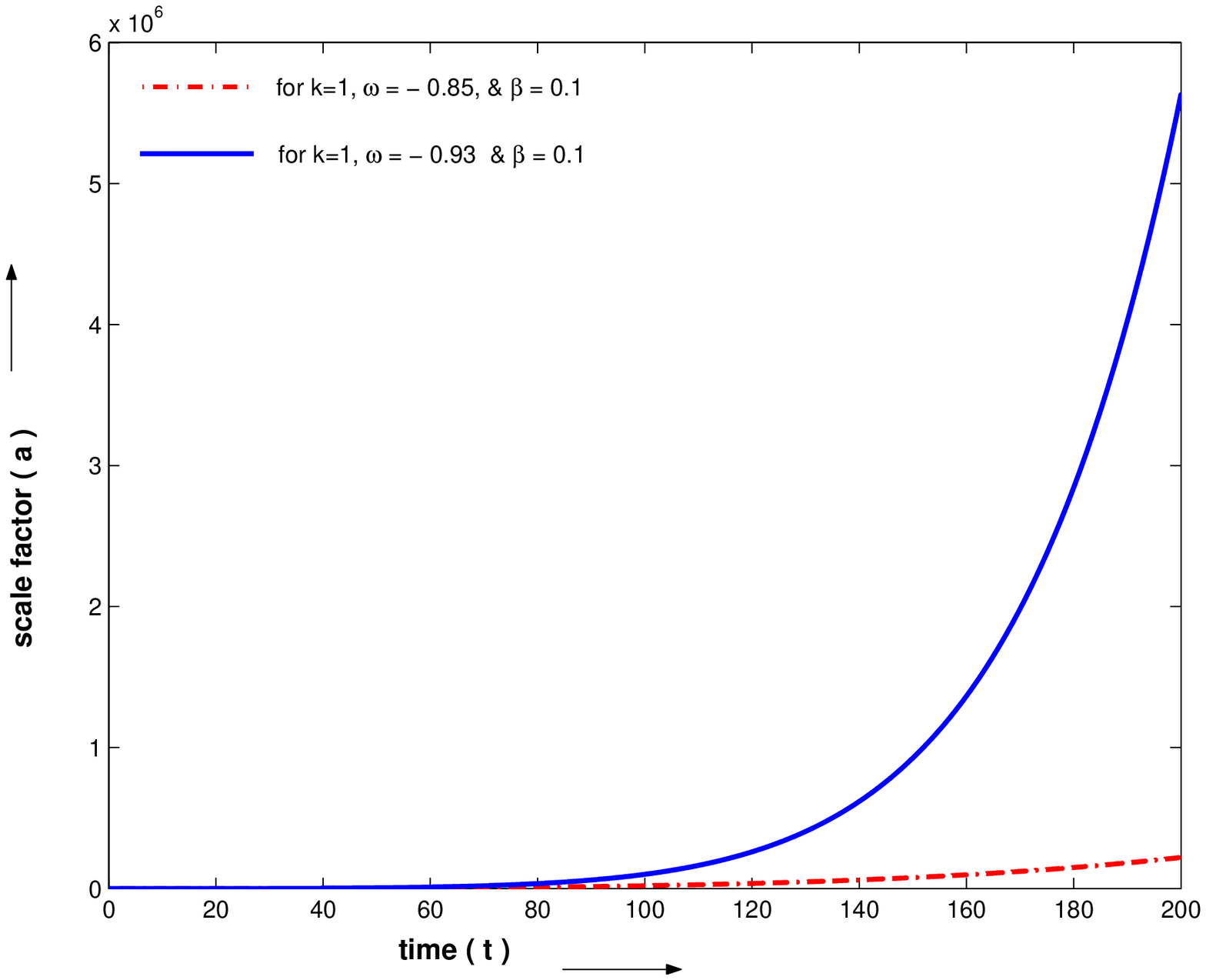,width=0.4\textwidth}
    \caption{Case-2.2 related to model $\Lambda = \beta (\ddot a/a)$
    shows variation of scale factor for $k=+1$.}
 \label{fig:2d}
\end{center}
\end{figure*}

\begin{figure*}
\begin{center}
\vspace{0.5cm} \psfig{file=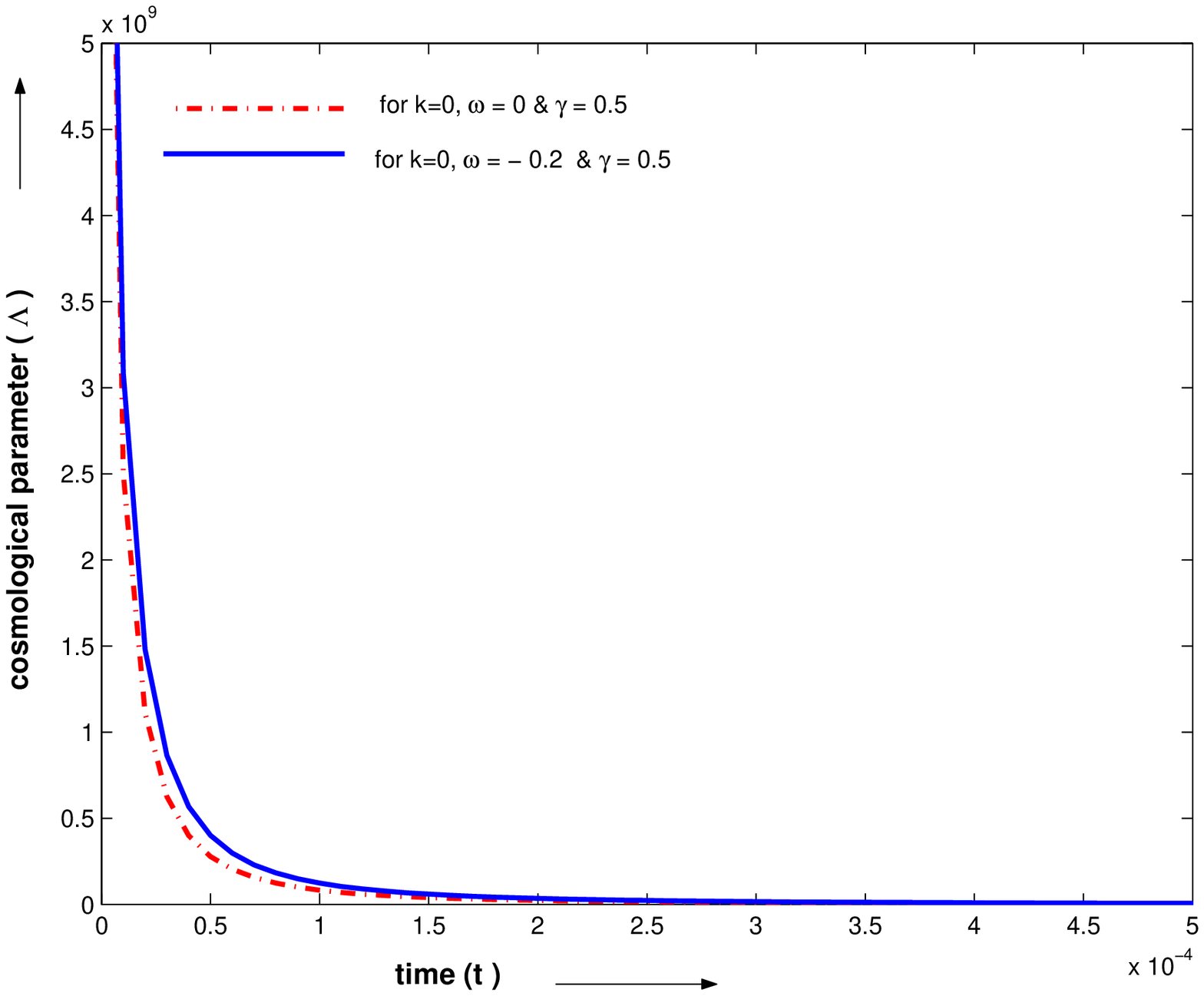,width=0.4\textwidth}
    \caption{Case-2.3 related to model $\Lambda = 8\pi G \gamma \rho$
    shows variation of cosmological parameter for $k=0$.}
 \label{fig:3a}
\end{center}
\end{figure*}

\begin{figure*}
\begin{center}
\vspace{0.5cm} \psfig{file=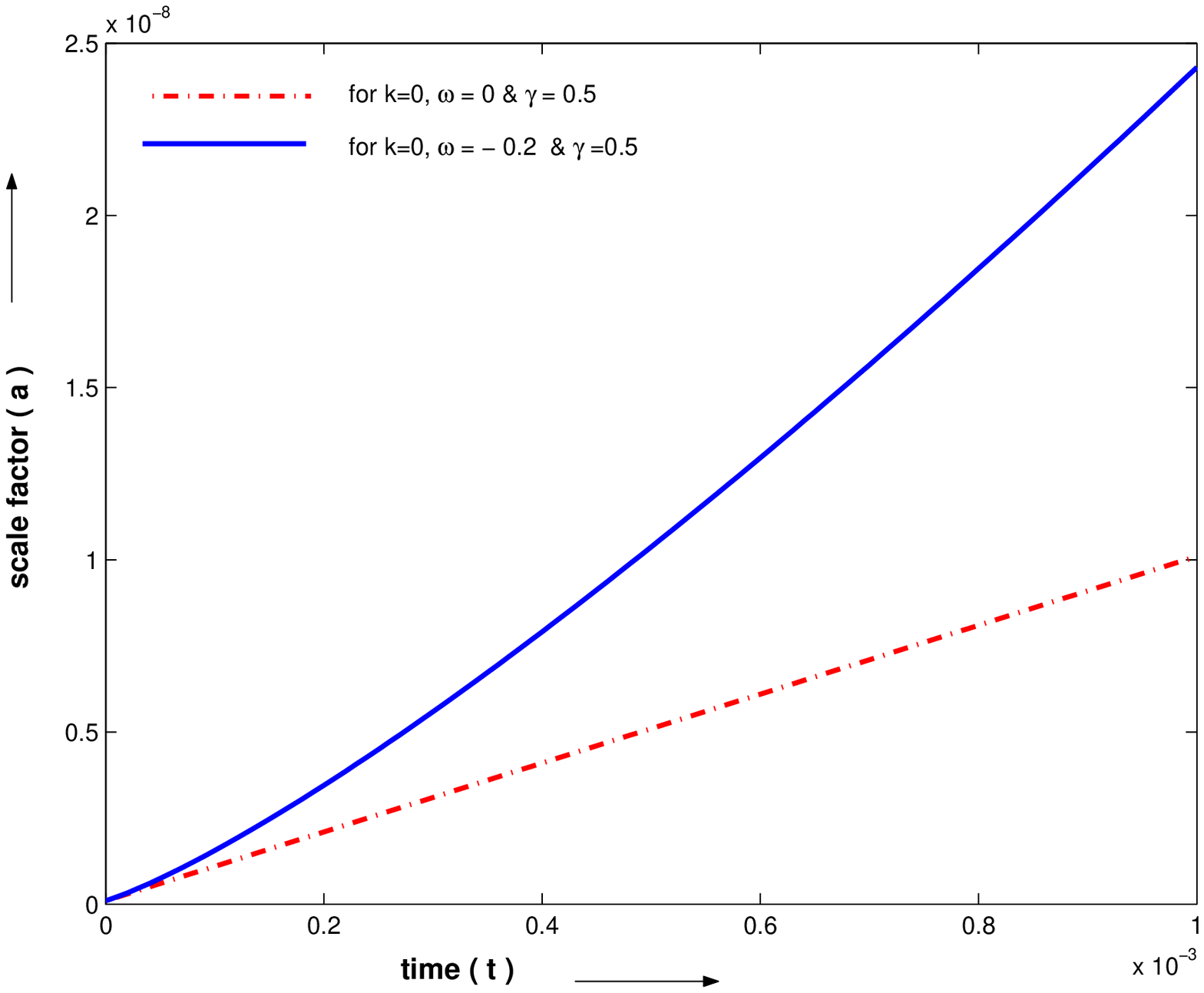,width=0.4\textwidth}
    \caption{Case-2.3 related to model $\Lambda = 8\pi G \gamma \rho$
    shows variation of scale factor for $k=0$.}
 \label{fig:3b}
\end{center}
\end{figure*}

\begin{figure*}
\begin{center}
\vspace{0.5cm} \psfig{file=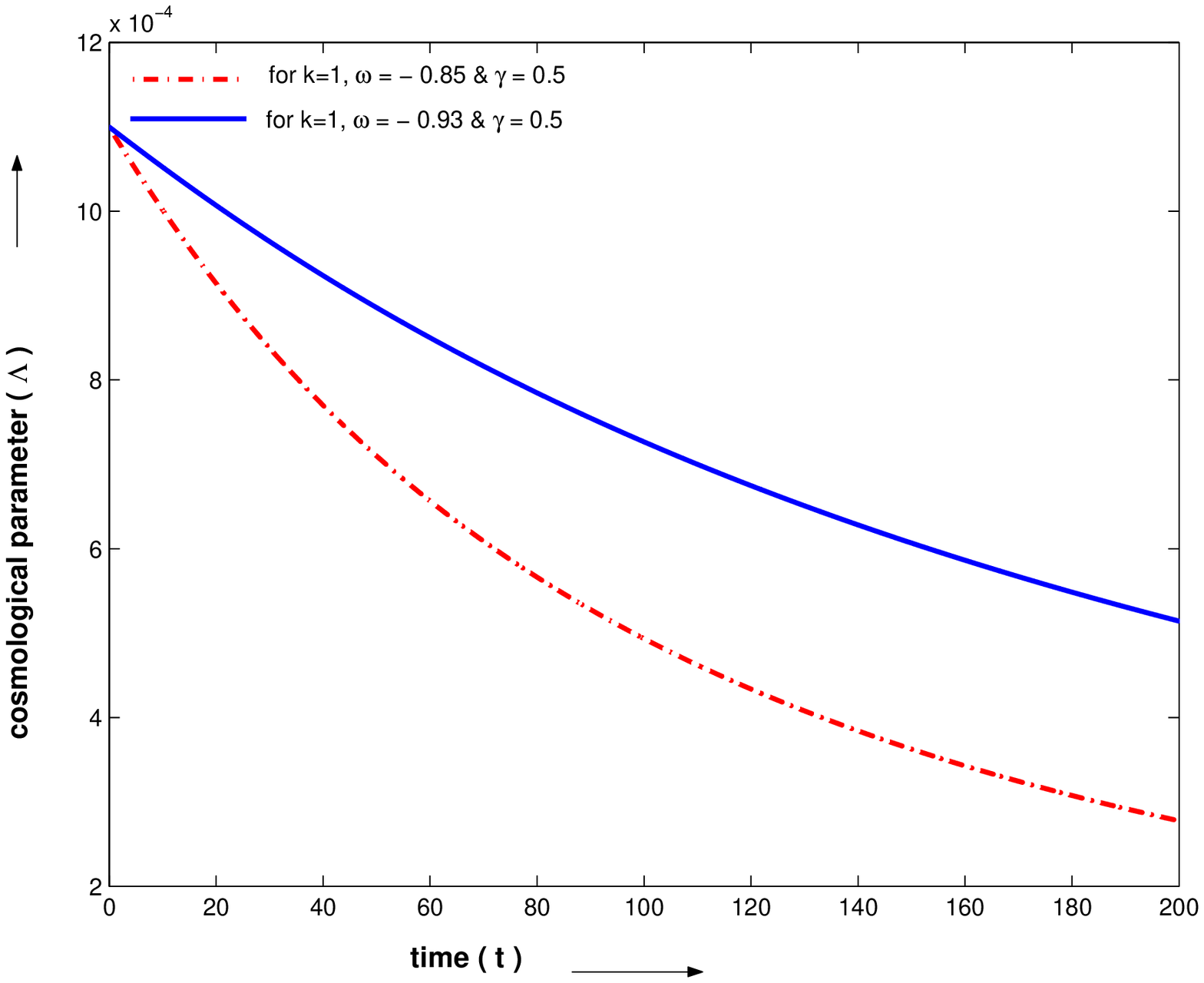,width=0.4\textwidth}
    \caption{Case-2.3 related to model $\Lambda = 8\pi G \gamma \rho$
    shows variation of cosmological parameter for $k=+1$.}
 \label{fig:3c}
\end{center}
\end{figure*}

\begin{figure*}
\begin{center}
\vspace{0.5cm} \psfig{file=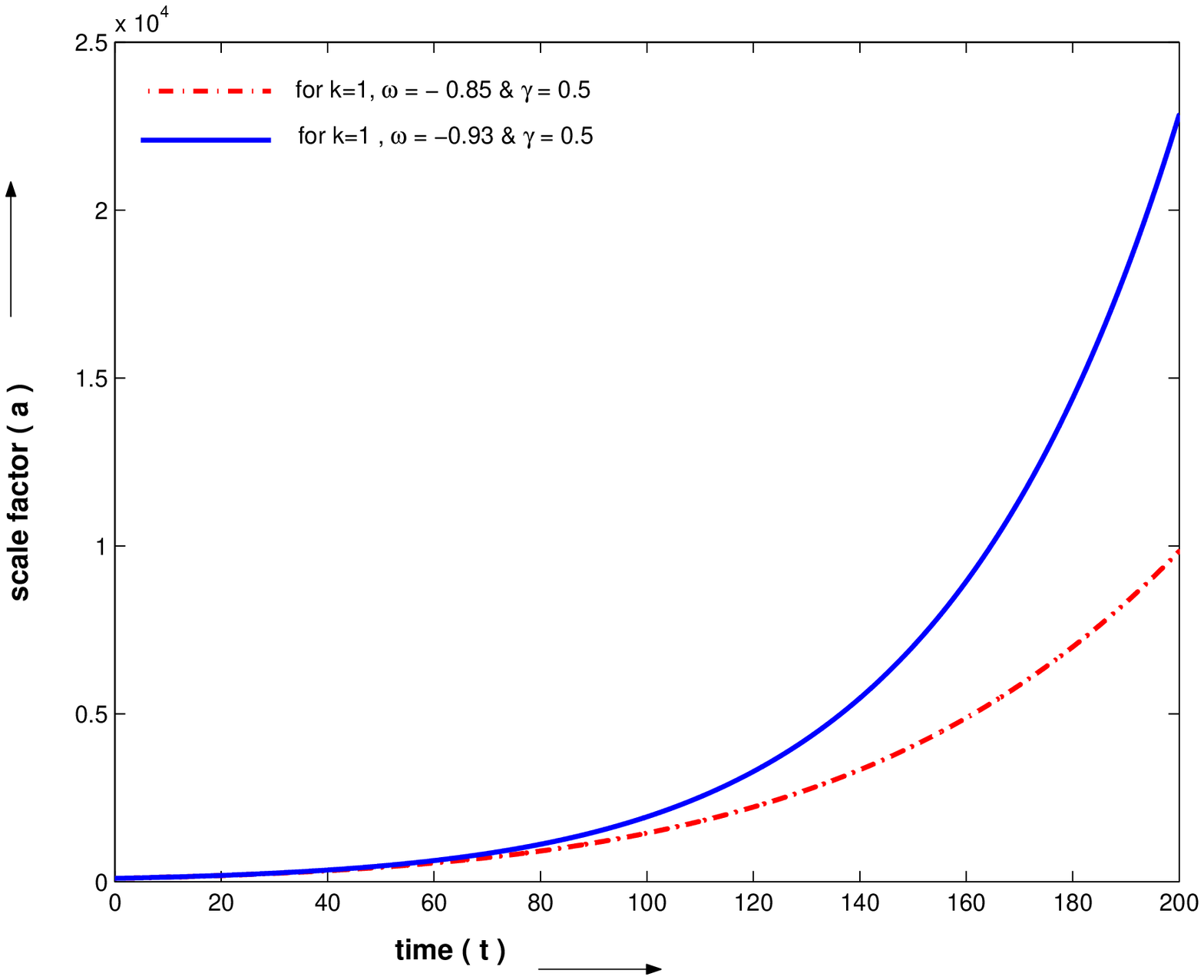,width=0.4\textwidth}
    \caption{Case-2.3 related to model $\Lambda = 8\pi G \gamma \rho$
    shows variation of scale factor for $k=+1$.}
 \label{fig:3d}
\end{center}
\end{figure*}

\end{document}